\documentclass[aps,pre,twocolumn,final,%
floatfix,showpacs,amsmath,amssymb,10pt]{revtex4}

\usepackage[T1]{fontenc}

\usepackage{times}

\usepackage{array}
\usepackage{graphicx}
\newcommand*\picturewidth{1.0\columnwidth}

\newcommand*\errorbars{Missing error bars are of the size of the symbols or smaller,  and omitted for legibility.\xspace}

\newcommand*{\legendlines}{%
(Parameter used: $\PairE=-2,\, \StackE=-1,\,\hmin=2$, 1000 samples.)
}

\newcommand*{\Base}[1][i]{\ensuremath{r_{#1}}}
\newcommand*{\cBase}[1][i]{\ensuremath{\bar{r}_{#1}}} %
\newcommand*{\Alphabet}{\ensuremath{\mathcal{A}}}
\newcommand*{\Letters}[1]{\texttt{#1}}
\newcommand*{\Sequence}{\ensuremath{\mathcal{R}}}
\newcommand*{\SubSequence}[1]{\ensuremath{\Sequence^{#1}}}
\newcommand*{\Structure}{\ensuremath{\mathcal{S}}}
\newcommand*{\SubStructure}[1]{\ensuremath{\Structure^{#1}}}
\newcommand*{\Chain}{\ensuremath{\mathcal{C}}}
\newcommand*{\Pair}{\wp}
\newcommand*{\UnDesignability}{\ensuremath{U}}
\newcommand*{\Designtime}{\ensuremath{T}}

\newcommand*{\SM}[1][i,j]{\ensuremath{S_{#1}}}

\newcommand*{\PairE}{\ensuremath{E_p}} %
\newcommand*{\StackE}{\ensuremath{E_s}}%
\newcommand*{\Emin}{\ensuremath{E_\text{min}}}%
\newcommand*{\Emax}{\ensuremath{E_\text{max}}}%
\newcommand*{\Extended}[1]{\ensuremath{#1^\ast}}%
\newcommand*{\et}{\Extended{e}_{\Sequence,\Structure}}
\newcommand*{\Joker}{\ensuremath{\ast}}
\newcommand*{\upperBound}[1]{{#1}^{\mathrm{B}}}
\newcommand*\hmin{\ensuremath{h_\text{min}}}
\newcommand*{\N}{\ensuremath{N}}
\newcommand*{\Nh}{\ensuremath{\hat N}}
\newcommand*\SN{\ensuremath{S}} %
\newcommand*\aSN{\ensuremath{s}} %

\newcommand*{\average}[1]{\ensuremath{\left\langle #1\right\rangle}}
\newcommand*{\define}{:=}

\newcommand*{\Tree}{\ensuremath{\mathcal{T}}}
\newcommand*{\proc}[1]{\textsc{#1}}

\usepackage{xspace}
\newcommand*{\eg}{e.g.,\xspace} %
\newcommand*{\ie}{i.e.,\xspace} %
\newcommand*{\refeq}[1]{Eq.\@ \eqref{#1}\xspace}
\newcommand*{\reffig}[1]{Fig.\@ \ref{#1}\xspace}
\newcommand*{\reftab}[1]{Tab.\@ \ref{#1}\xspace}
\newcommand*{\refsec}[1]{Sec.\@ \ref{#1}\xspace}

\newcommand*{\etal}{\textit{et al.\xspace}}

\begin{document}

\author{Bernd Burghardt}
\email{burghard@physik.uni-goe.de}
\author{Alexander K. Hartmann}
\email{hartmann@physik.uni-goe.de}
\affiliation{Institut f{\"u}r Theoretische Physik, Universit{\"a}t G{\"o}ttingen,
  Friedrich-Hund-Platz 1, D--37077 G{\"o}ttingen, Germany}

\title{RNA secondary structure design}
\date{\today}

\begin{abstract}
We consider the inverse-folding problem for RNA secondary structures: for a
given (pseudo-knot-free) secondary structure find a sequence that has that
structure as its ground state. If such a sequence exists, the
  structure is called designable.
We implemented a branch-and-bound algorithm that is able to do an exhaustive 
search within the sequence space, \ie gives an exact answer
  whether such a sequence exists.
The bound required by the  branch-and-bound algorithm are calculated by a
dynamic programming algorithm. 
We consider different alphabet sizes and an ensemble of random structures,
which we want to design. 
We find that for two letters almost none of these structures are designable.
The designability improves for the three-letter case, but still a
significant fraction of structures is undesignable. 
This changes when we look at the natural four-letter case with two pairs of
complementary bases: undesignable structures are the exception, although they still
exist. Finally, we also study the relation between
  designability and the algorithmic complexity of the branch-and-bound
  algorithm. Within the ensemble of structures, 
a high average degree of undesignability is correlated to a long
  time to prove that a given structure is (un-)designable. In the
  four-letter case, where the designability is high everywhere, the algorithmic
  complexity is highest in the region of naturally occurring RNA.
\end{abstract}


\maketitle

\section{Introduction}
RNA plays an important role in the biochemistry of all living
systems~\cite{GCA99,Hig00}.  Similar to the DNA, it is a linear chain-molecule
build from four types of bases---\ie adenine (\Letters A), cytosine
(\Letters C), guanine (\Letters G), 
and uracil (\Letters U).  
It does not only transmit pure genetic information, but, \eg works as a
catalyst, for example in the ribosome. 
While for the former only the primary structure---\ie the sequence of
the bases---is relevant, for the latter the kind of higher order
structures---\ie secondary and tertiary structures, is essential for
its function.
We exemplary mention the following
three examples: 
i)  For successful protein synthesis three-dimensional structures of
rRNA \cite{Nol84,GN97}  %
 and tRNA \cite{KSQ*74} %
molecules are inevitable. 
ii) The catalytic properties of ribozymes depend on their
three-dimensional structures\cite{KGZ*82}. %
iii) The function of the internal ribosome entry site (IRES) of
picornaviruses which directs binding of ribosomal subunits and
cellular proteins in order to accomplish translation initiation, is
based on higher order structures\cite{JSB03}.

Like in the double helix of the DNA, complementary bases  within RNA
  molecules can build
hydrogen bonds between each other. As opposed to DNA, where the bonds are built
between two different strands, in RNA bonds are formed
between bases of the same RNA
strand. The {\em secondary structure} is the
 information, which bases of the strand are paired, 
while the spatial structure is called the {\em tertiary
structure}. 
The tertiary structure is stabilized by a much weaker interaction than the
secondary structure. This leads to a separation of energy scales between
secondary and tertiary structure, and gives the justification  to neglect the
latter in many cases to obtain a first fundamental 
understanding of the behavior  of RNA~\cite{BH99}. Therefore, 
although the
tertiary structure is important often for an RNA's functionality, it is
sufficient that we deal here with the secondary structure only.

One crucial point for the  calculation of the secondary 
structure is the energy model, which is applied: 
On the one hand, if one aims to get minimum structures close to the
experimentally observed one, one uses energy models that take into account
many different structural elements
~\cite{Zuk89,McC90,HFS*94,LZP99}, \eg hair pin loops or
  bulges, each being described by a different set of experimentally
  obtained parameters. 
On the other hand, if one is interested in the qualitative behavior, one uses
models as simple as possible while conserving
the general behavior, \eg in the simplest case 
a model which exhibits only one
kind of base~\cite{LB04} or models where the energies depend
 only on the number and on 
the type of paired bases~\cite{Hig96,BH02,MPR02,PPR00}. 
Here we will consider only models with the latter kind of interaction energy. 

The standard procedure when dealing with RNA secondary 
structures is that one starts with a given
sequence and calculates, \eg the ground-state structure in which the RNA will
fold for low temperatures. 
In this paper we look at the inverse problem: For a given
secondary structure, does a sequence exist 
that has the given structure as its
ground state?  If this is the case, we call the structure {\em
  designable}. We answer this question for different alphabet
sizes, \ie different numbers of complementary bases. 
As an ensemble of structures we choose a set of random 
structures of given length and ask
how large is the fraction designable structures.
In a related study 
Mukhopadhyay \etal \cite{MET*03} also considered different alphabet sizes, but
they determined   
for a ground-state structure of a given sequence, by using a
  probabilistic algorithm, \ie  approximately, how many different 
{\em other} sequences have this structure as a ground state. 
Hence, by definition, all structures encountered are designable.
In contrast, we generate structures randomly from scratch,
  and determine whether there is at least 
one sequence  that has this structure as a ground state. 
Hence, we can generate structures, which might not be designable at all.
The basic idea behind this approach is that nature needs as many different
structures as possible to perform many different tasks, and, as it
turns out, a minimum number of four letters is necessary for this. 
Furthermore, we use an exact branch-and-bound 
algorithm to verify (un-)designability.
In another previous work Hofacker
\etal \cite{HFS*94} (with improvements by \cite{AFH*04}) looked at the
same question  whether a given structure is designable.
In contrast to our  work, they used only
 a probabilistic approach, hence in some cases solutions may have been
  missed. Furthermore, 
they studied a very restricted ensemble of structures, where the 
structures are assembled from substructures found in nature already,
  which implies by definition a high degree of designability.
Also they did not study the dependence on the alphabet
  size. Another difference of our work to previous publications is that
 we also study the relation between the
  designability and the algorithmic complexity, \ie the running time
  of our exact algorithm.

The paper is organized as follows. In section \refsec{sec:model}, we define
our model---\ie we formally define secondary structures and introduce
our energy model and state the design problem. 
In \refsec{sec:algorithm}, we explain how to calculate a bound for the ground
state with a dynamic programming algorithm and how to solve the design problem
with a branch-and-bound algorithm augmented with a randomized algorithm.
We also present thoroughly in \refsec{sec:structure_generation} how we generate
the ensemble of random structures. 
Finally, in \refsec{sec:results} we show the  result of 
our numerical studies.

\section{The secondary structure model  and design problem}
\label{sec:model}

\subsection{RNA secondary structure model}
\label{sec:structure_model}
Because RNA molecules are linear chains of bases, they  can be described as a
(quenched) sequence 
$\Sequence=(\Base[i])_{i=1,\dots,L}$ of bases
$\Base[i]\in\Alphabet$. We denote by $L$ the length of 
the sequence and 
$\Alphabet$ is the alphabet, which  contains 
the underlying base types that build the RNA
sequence. Typically $\Alphabet=\{\Letters{A,C,G,U}\}$ is used,
but we also consider here alphabets with two and three letters. 
Within this single stranded molecule some bases can pair and build a secondary
structure. 
The Watson-Crick base pairs --- \ie \Letters A-\Letters U 
and \Letters C-\Letters G --- have the strongest
affinity to each other, they are also called
   complementary base pairs. 
Each base can be paired at most once.
For a given sequence $\Sequence$ of bases the secondary structure can be
described by a set $\Structure$ of pairs $(i,j)$ (with the convention $1\leq
i<j\leq L$), meaning that bases \Base\ and $\Base[j]$ are paired. 
For convenience of notation we further define a Matrix $(\SM)_{i,j=1,\dots,L}$
with $\SM=1$ if $(i,j)\in\Structure$, and $\SM=0$ otherwise.  
Two restriction are used: 
\begin{enumerate}
\item \textsl{[non-crossing condition]}
Here we exclude so called \emph{pseudo knots}, that means, for any
$(i,j),(i',j')\in\Structure$, either $i<j<i'<j'$ or $i<i'<j'<j$ must
hold---\ie we follow the notion of pseudo knots being more an element of the
tertiary structure \cite{TB99}.

\item\textsl{[min-distance condition]}
 Between two paired bases a minimum distance is required: $|j-i|\geq \hmin$ is
required, due to the bending rigidity of the
molecule. Our main results below will be for $\hmin=2$, but
for comparison we discuss the unphysical case $\hmin=1$ as well.
Larger---and more realistic---$\hmin$ values do not change the
  qualitative results compared to the  $\hmin=2$ case, but are computationally
more demanding.
\end{enumerate}

In the following we assume that each 
structure \Structure\ 'fits' to all considered sequences
\Sequence---\ie for all pairs $(i,j)\in\Structure$ the indices $i$ and $j$ are
smaller or equal to  the length $L$ of the sequence ($1\leq i,j\leq L$).
By $\SubStructure{m,n}$ we denote a \emph{substructure} of
$S$ between the $m$'th and $n$'th letter, \ie
$\SubStructure{m,n}\define \left\{(i,j)\in\Structure\mid m\leq i<j\leq
  n\right\}$. 
Similar, a \emph{subsequence} between the $m$'th and $n$'th letter is denoted
by $\SubSequence{m,n}=(\Base[i])_{i=m,\dots,n}$.

\subsection{Energy models}
\label{sec:energy_models}
In this section we define an energy model, which assigns every 
secondary structure \Structure\ belonging to a sequence \Sequence\ an
energy $E(\Structure, \Sequence)$.
For a given sequence \Sequence\ the minimum
$E(\Sequence)=\min_{\Structure}E(\Structure, \Sequence)$ 
is the ground-state energy of the sequence \Sequence.  

Motivated by the observation that the secondary structure is due to building
of numerous base pairs where every pair of bases is formed via hydrogen
bonds, one assigns each pair $(i,j)$ a
certain energy $e(\Base,\Base[j])$ depending only on the kind of bases.
The total energy is the sum over all pairs
\begin{equation}
  \label{eq:paired_energy_model}
\PairE(\Structure, \Sequence)=\sum_{(i,j)\in\Structure}e(\Base,\Base[j])\,,
\end{equation}
\eg by choosing $e(\Base[], \Base[]')=+\infty$ for non-complementary bases
\Base[], $\Base[]'$ pairings of this kind are suppressed.
In our numerical studies we restrict our self to  the energy model
\begin{equation}
\label{eq:pair_energy}
 e(\Base[],\Base[]')=
 \begin{cases}
   \PairE& \text{if $r$ and $r'$ are compl.\ bases}\\
   +\infty& \text{otherwise}
 \end{cases}
\end{equation}
with a pair energy $\PairE\leq0$ independent of the kind of bases.

Another possible model is to assign an energy \StackE\ to a pair
$(i,j)\in\Structure$  iff also $(i+1,j-1)\in\Structure$. This \emph{stacking
  energy} can be motivated by the fact that a single pairing gives some gain
in the binding energy, but also reduces the entropy of the molecule, because
through this additional binding it looses some flexibility.
 Formally the total
energy of a structure can be written as
\begin{equation}
  \label{eq:stacked_energy_model}
  \StackE(\Structure, \Sequence)=
  \begin{cases}
    \sum_{(i,j)\in\Structure}\StackE\SM[i+1,j-1] & \text{if all } (i,j)\in\Structure:\\
    &\Base[i],\Base[j] \text{are compl.\ bases}\\
    +\infty & \text{otherwise}
  \end{cases}
\end{equation}
 Real RNAs cannot be described by just one energy
parameter, because the free energy depends  
on the type and the size of the structural elements,
 \eg hair pin loops. 
Here, we examine the sum of both models---stacking energy and pair energy---  
\begin{equation}
  \label{eq:energy_model}
  E(\Structure, \Sequence)\define \PairE(\Structure, \Sequence)+ \StackE(\Structure, \Sequence),
\end{equation}
where the parameters \StackE\ and $e(\Base[],\Base[]')$ can be freely adjusted,
including both models discussed above.
For real RNA both parameters, $\PairE$ and $\StackE$, are of the same order of
magnitude, namely about $1\dots10\mathrm{kcal}\,\mathrm{mol}^{-1}$
\cite{BTT99,Hig93,Zuk89}, therefore we choose $\PairE=-2$ and $\StackE=-1$ in
our simulations.

A sequence \Sequence\ is said to be \emph{compatible} with a structure
\Structure, if $e(\Base, \Base[j])\leq0$ for all $(i,j)\in\Structure$.

Further, we define for a structure \Structure\ (independent of
\Sequence) the energy  
\begin{equation}
  \label{eq:structure_energy}
  E(\Structure)\define \Emin\,|\Structure|+\sum_{(i,j)\in\Structure}\StackE\SM[i+1,j-1]\,,
\end{equation}
with $\Emin=\min_{r,r'\in\Alphabet}e(r,r')$.
For the energy model of \refeq{eq:pair_energy} it is $\Emin=\PairE$.
Thus, $E(\Structure)$ is 
a lower bound of $E(\Structure, \Sequence)$ for any \Sequence.

\subsection{Designing RNA Secondary Structure}
\label{sec:design_problem}

The energy model (\ref{eq:energy_model}) has been previously studied
\cite{BH05}, in the standard way, \ie by calculating ground states
for given sequences.
In this paper we take, as already mentioned in the introduction,
 a different point of view: we choose a random
\emph{structure} \Structure\ and ask, whether there exists
any sequence \Sequence\ that has this
structure as its ground state. 

The \emph{design problem} can be more formally stated as following: 
For a given structure \Structure\ find a sequence \Sequence\ such that
$E(\Structure, \Sequence) = E(\Sequence)$ holds. 
If such a sequence exists, the structure \Structure\ is called
{designable}. 
However, we do not require that \Structure\ is  the unique ground state of
this sequence, since this issue has been addressed previously
  \cite{MET*03}.

\begin{figure}[htbp]
\includegraphics*[width=\picturewidth]{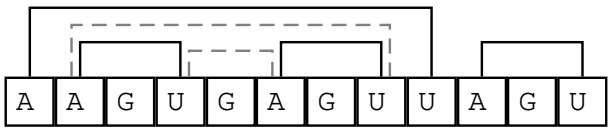}
  \caption{In the case $\StackE=0$ the structure can be easily designed, \eg by
    building $(\Letters A,\Letters U)$-pairs for the paired bases, and
    assigning $c$ to the   unpaired bases. However, this is not
    necessarily  a solution for the $\StackE<0$ case: in this example
    two pairs could be re-paired (dashed lines) giving a lower overall
    energy.   \label{fig:designable}}   
\end{figure}

The design problem for an energy model \emph{without}
 stacking energy, \ie which exhibits only a pair energy
according to \refeq{eq:pair_energy}, can be solved
easily as follows (\reffig{fig:designable}):
assign to any pair $(i,j)\in\Structure$ the letters 
\Letters A at position $i$ and \Letters U at position  $i$, and for every
unpaired position a base of type \Letters G (in the two letter case
use \Letters A again). There 
are exactly $|\Structure|$ pairs of bases therefore the 
ground-state energy
can not be below $\PairE\,|\Structure |$, which is just the ground-state energy
of the structure \Structure.

For the case $\StackE\lneqq0$ this construction scheme might fail as one can
see in the example shown in \reffig{fig:designable}: 
re-grouping of the enclosed base pairs leads to the formation
  of two adjacent pairs, \ie a stack of size two. This results in
an energy of
the re-grouped 
structure below the energy of the given structure, hence the
  given structure is
not a ground state of the given sequence. Nevertheless, the
  structure shown in the example is in fact designable, the slightly
    modified sequence---position 2 and 4 are swapped---
 \Letters{AUGAGAGUUAGU} has the given structure as a ground state.

\begin{figure}[htbp]
\includegraphics*[width=\picturewidth]{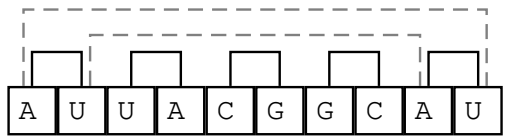}
  \caption{In the case of $\hmin=1$ and $\StackE<0$ this is an example of an
    undesignable   structure. There is only a
finite number of  different 2-tuples $(\Base[1],\Base[2])$.
 Whenever there  
are more than  this number of neighboring pairs paired in a structure, at
least two of  them must be of the same kind, \eg
$(\Letters{A},\Letters{U})$, this  
two can be re-paired {(dashed  lines)}
 gaining some {stacking} energy, 
rendering the structure undesignable.   \label{fig:hmin_counter_example}
}  
\end{figure}

The case $\hmin=1$, \ie neighboring bases can be paired, is of little
interest: both, from the physical point of view---the RNA molecule cannot be
bent arbitrarily strong---as well as from the design
problems point of view. 
As an undesignable example look at the structure sketched in figure
\reffig{fig:hmin_counter_example}: for any alphabet size there is only a
finite number of different 2-tuples $(\Base[1],\Base[2])$, whenever there
are more than this number of neighboring pairs paired in a structure, at
least two of them must be of the same kind---\eg
$(\Letters{A},\Letters{U})$---this two can be re-paired and 
gaining some stacking energy, rendering the structure
undesignable. 

\section{Algorithms}
\label{sec:algorithm}
In principle the design problem can be solved by calculating the ground state
energy $E(\Sequence)$ of every compatible sequence \Sequence\ and testing whether this is
equal to $E(\Structure, \Sequence)$, but, because the number of sequences
growth exponentially with the sequence size $L$ (roughly as
$|\Alphabet|^{L-|\Structure|}$), this is impractical. 

Therefore we use a branch-and-bound algorithm, where one tries to find an
upper bound $\upperBound E(Q)\define \max_{\Sequence\in Q} E(\Sequence)$ for
the ground-state energies for a (large) set $Q$ of sequences compatible with
the structure \Structure. If this bound is below the energy $E(\Structure)$ of
the structure---\ie $\upperBound E(Q)<E(\Structure)$---then none of the
sequences in $Q$  can be a solution of the design problem.

Here, we consider in particular sets of sequences, where at some
positions all sequences of the set have the same letter (but possible different
ones for the different positions), and where for all other positions
all possible combinations of letters occur, which are compatible with
the sequence. Hence, these positions can be described by a \emph{joker}
letter. For a more formal definition of $Q$, see below. In
\refsec{sec:nussinov_algorithm} an algorithm is explained, which 
calculates an
upper bound for the ground-state energy of such sequences.

This algorithm is used within
the bound step of the branch-and-bound algorithm, which is
 explained in \refsec{sec:branchandbound}.

\subsection{Calculating a bound for the ground-state energy}
\label{sec:nussinov_algorithm}
In this section we introduce a modification of the algorithm presented in
Ref.\ \onlinecite{BH05} which allows us
to calculate an upper bound for the ground-state energy of
sequence, where some bases are still unassigned, \ie
  represented by the joker letter.

Thus,
for a formal description of the algorithm we extend the \Alphabet\ by the
joker-letter \Joker, 
where \Joker\ represents any letter in the original alphabet. 
Note that \Joker\ is complementary to any $r\in\Alphabet$.
  The new  alphabet is denoted by
$\Extended\Alphabet\define\Alphabet\cup\left\{\Joker\right\}$.  
Sequences $\Extended\Sequence=(\Extended r_i)_{i=1\dots L}$,
$\Extended r_i\in\Extended\Alphabet$, 
over this extended alphabet $\Extended\Alphabet$, we
  call $\Extended\Sequence$ a {\em generalized sequence},
represent a set $Q$ of sequences over the original
\Alphabet: $Q=\left\{(r_i)_{i=1,\dots,L}| r_i\in\Alphabet,
  r_i=\Extended r_i \text{ if }\Extended r_i\in\Alphabet\right\}$. 
For a given structure \Structure\ and a generalized sequence 
$\Extended\Sequence$,  the scheme explained in the following can be
used to calculate the a bound for the ground-state
energy. Note that for a sequence
without a \Joker--letter this bound is equal to the ground-state
energy.

We start the explanation of the algorithm by considering the
contribution to the bound arising from a single  pair $(i,j)$. 
If the letters in the sequence are fixed,
\ie $\Base,\Base[j]\in\Alphabet$, then the energy contribution 
 is simply $ e(\Base,\Base[j])$, since there is
no choice. If at least one
of the two letters is the joker letter $\Joker$, then we have
different choices. 
First, if $(i,j)\in\Structure$, then the energy
contribution must be negative, because otherwise, since we are
considering ground states, bases $i$ and $j$ would not be paired
leading to an energy contribution zero. On the other hand, we are
looking for an maximum over all sequences described by the generalized
$\Extended\Sequence$,  hence we have to take the maximum over all
possible negative contributions, either over all possible combinations
of two letters (two $\Joker$ symbols), or, over all possible letters
at the one position with a $\Joker$ symbol. 
Second, if $(i,j)\not\in\Structure$,
then the energy contribution should be positive if bases $i$,$j$ are
paired nevertheless, such that within the ground-state calculation,
automatically the case is selected where bases $i,j$ are not paired. We
assume that for all possible cases with one or two $\Joker$ symbols,
always combinations of letters are available, such that the pair
energy is positive. Since in this case, the ground-state requirement
will automatically disregard the pair $(i,j)$, instead of maximizing
over all energies, we can simply assume the energy contribution
$+\infty$ here.
This leads to the energy contribution $\et(i,\,j)$ for a pair $(i,j)$
which depends on the given
generalized sequence \Extended\Sequence\ and the given structure 
$\Structure$: 
\begin{align}
  \label{eq:energy_definition}
  \et(i,\,j)&=
  \begin{cases}
    e(\Base,\Base[j]) & \text{if } \Base,\Base[j]\in\Alphabet\wedge |i-j|\geq\hmin\\
    \Emax^{\Joker,\Joker}%
    & \text{if } \Base=\Joker,\Base[j]=\Joker,
    (i,j)\in\Structure\\
    \Emax^{r_i,\Joker}%
    & \text{if } \Base\in\Alphabet,\Base[j]=\Joker,
    (i,j)\in\Structure\\
    \Emax^{\Joker,r_j}%
    & \text{if } \Base=\Joker,\Base[j]\in\Alphabet,
    (i,j)\in\Structure\\
    +\infty           & \text{else}
  \end{cases}\\
\intertext{with the largest possible negative pair energies}
  \Emax^{\Joker,\Joker}%
  &\define \max\left\{e(r,r')<0|r,r'\in\Alphabet\right\} \nonumber\\
  \Emax^{r,\Joker}%
  &\define \max\left\{e(r,r')<0|r'\in\Alphabet\right\}\label{eq:joker_energies} \\
  \Emax^{\Joker,r'}%
  &\define \max\left\{e(r,r')<0|r\in\Alphabet\right\} \nonumber 
\end{align}
and for the maximum of the empty set: $\max\emptyset\define-\infty$.
For alphabets, where each base has a complementary base, \eg the
  two- and four-letter cases discussed below,
  with the energy $e(r,r')$ from \refeq{eq:pair_energy} $ \et$ has the form
\begin{align}
  \label{eq:energy_definitionB}
  \et(i,\,j)&=
  \begin{cases}
    e(\Base,\Base[j]) & \text{if } \Base,\Base[j]\in\Alphabet\\
    \PairE  
    & \text{if } \Base=\Joker\vee\Base[j]=\Joker,
    (i,j)\in\Structure\\
    +\infty           & \text{else}
  \end{cases}
\end{align}
For alphabets with letters that have no complementary counterpart, \eg letter
\Letters{G} in the three-letter alphabet of \refsec{sec:three_letter}, the
sets in \refeq{eq:joker_energies} might be empty leading to an energy
contribution $-\infty$, 
\ie resulting in an upper bound $\upperBound{E}(\Extended\Sequence)=-\infty$.
In our implementation of the algorithm we do not consider (generalized)
sequences, where at a position of a paired base such a letter appears, because
this would lead do non-compatible sequences.
Note that for the case that also the pair $(i-1,j+1)$ is present,
additionally to   $\et(i,\,j)$ a
stacking-energy contribution $\StackE$ arises. This is handled by the following
recursive equations, which perform the ground-state calculation. They are
slightly modified compared to Ref.\
\onlinecite{BH05}. We denote by $\N_{i,j}$ the maximum 
ground-state energy over  the set of compatible subsequences given by 
the generalized subsequence
 $\Extended r_i, \Extended r_{i+1}, \ldots, \Extended r_{j-1},
 \Extended r_j$. 
$\Nh_{i,j}$ is defined in the same way, only that
 additionally it is assumed that letters $\Extended r_{i-1}$ and
 $\Extended r_{j+1}$ are paired, which leads simply to an additional
 stacking-energy contribution. The basic idea is that for the ground
state of subsequence $\Extended r_i,\ldots,  \Extended r_j$ either
the last letter $j$ is not paired, or it is paired to another letter
$k\in\{i,i+1,\ldots,j-1\}$ (the requirement $j-i\geq\hmin$ is treated through
energy $\et(i,\,j)$). The ground state is the minimum over all these
cases, where in each case, due to the exclusion of pseudo knots,
 the ground-state calculation
decomposed into  the calculation for shorter subsequences.  
The recursion equations for  $\N_{i,j}$
and $\Nh_{i,j}$ read as follows. 
  \begin{align}
  \label{eq:nuss_recursion}
  \N_{i,j}=& \min\left\{\N_{i,j-1},\right.\nonumber\\
& 
 \left.\min_{k=i}^{j-1}\left[\N_{i,k-1}+\et(k,
   j)+\Nh_{k+1,j-1}\right]\right\}\nonumber\\
& \text{for }j-i>0
\nonumber\\
       \Nh_{i,j}=&\min\left\{\N_{i,j-1}, \et(i, j)+\StackE+\Nh_{i+1,j-1},\right.
       \\
     &  \left.
     \min_{k=i+1}^{j-1}\left[  
     \N_{i,k-1}+ \et(k, j)+\Nh_{k+1,j-1}   \right]\right\}\nonumber\\
& \text{for }j-i>0 \nonumber\\
     & \N_{i,j}= \Nh_{i,j}=0\quad\text{for } j-i\le 0 \nonumber
\end{align}

The values of $\N_{i,j}$ and $\Nh_{i,j}$ are calculated
  ``bottom up'', \ie in a {\em dynamic programming} fashion,
starting at small values of $j-i$ until one arrives at $j-i=L-1$.
The wanted bound is $\upperBound{E}(\Extended\Sequence)=\N_{1,L}$, and within
our energy model this bound is never larger than $E(\Structure)$.
  In general, $\N_{i,j}$ is the bound for the ground-state energy of the subsequence $(\Extended r_{k})_{k=i,\dots,j}$.

It is worthwhile to note that it is not necessary to recalculate the whole
matrix $(\N_{i,j})_{1\leq i\leq j\leq L}$ if only one letter in
$\Extended\Sequence$ has been changed, \eg if base $\Base[k]$ has been
modified
  this  only influences subsequences which contain this base, therefore
it suffices to recalculate all $\N_{i,j}$ and $\Nh_{i,j}$ with $i\leq k\leq j$.
This reduces the numerical effort for calculating $\N_{1,L}$, but it is still
of order $\mathcal{O}(L^3)$.

\subsection{Algorithms for solving the design problem}
\label{sec:design_algorithms}

In this section we describe two algorithms, which we used to solve the
design problem stated above.
The first one is a deterministic, \ie it  guarantees to either
successfully find a solution or to prove that no solution exists. 
For this the algorithm has to consider 
exponentially (in the length $L$) many sequences. 
In the case that the problem has a solution a randomized algorithm is
often faster in finding a solution, therefore we also implemented such
an algorithm \cite{SFS*94,AFH*04}, and combined both algorithms.

\subsubsection{Branch-and-Bound algorithm}
\label{sec:branchandbound}

Our deterministic algorithm follows the Branch-and-Bound 
approach 
(\eg in Ref.\ \onlinecite[pp.~499]{KV02}). Here, it finds
a sequence \Sequence---if
such a sequence exists---that has the \Structure\ as one ground-state.

The idea of the algorithm is that it constructs a tree, 
where each node represents a generalized sequence
  \Extended\Sequence,  \ie a set  $Q$ of sequences, 
and all children of a node represent a partition of $Q$.
The root node
stands for the set of all sequences of length $L$, \ie which
is described  by the generalized sequence 
$(\Extended\Base)_{i=1,\dots,L},\,\Extended\Base=\Joker$. 
For every node $(\Extended\Base)$ in the tree
with at least one $\Extended{\Base[j]}=\Joker$ its children are constructed by replacing
$\Extended{\Base[j]}$ with one letter from \Alphabet.
Sequences with no $\Joker$-letters are the leaf nodes of the tree (sets with
exactly one element/sequence).

\begin{figure}[htbp]
  \includegraphics*[width=\picturewidth]{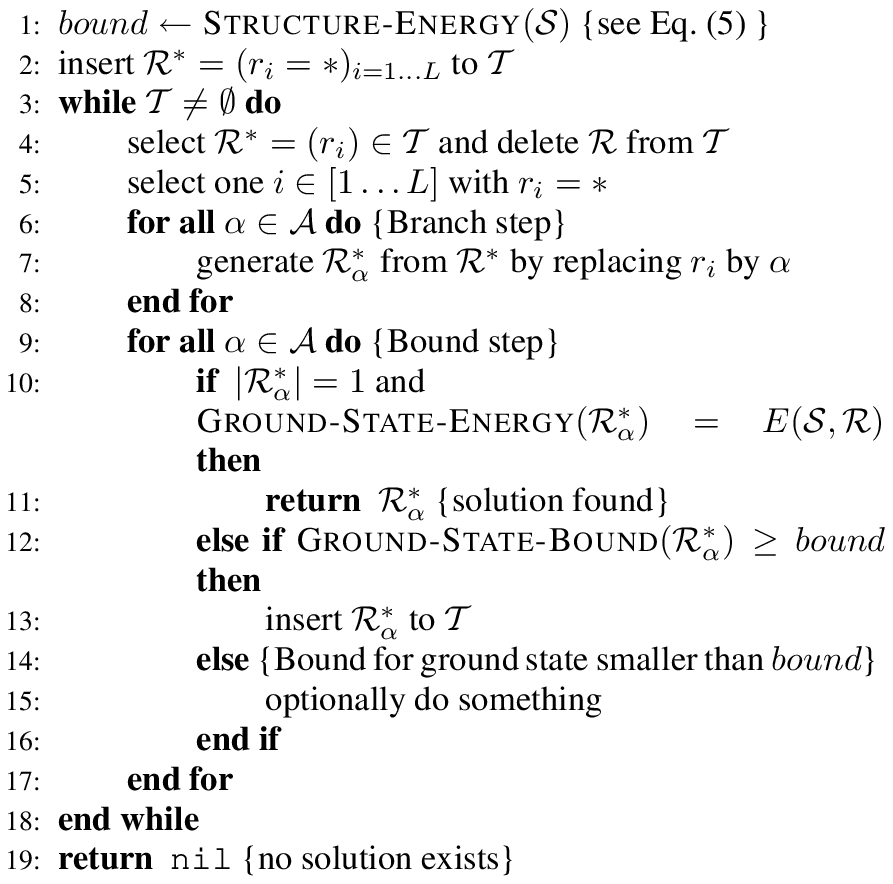}
 \caption{Pseudo-code of the branch-and-bound algorithm. 
    In line 15 the algorithm can be augmented, \eg with an
    randomized algorithm--see \refsec{sec:randomized_algo}.
  \label{fig:branch_and_bound_algorithm}
  }
\end{figure}

In  \reffig{fig:branch_and_bound_algorithm}, a pseudo code of
  the algorithm is shown. There, $\Tree$ contains all nodes of
  the tree which have not been treated yet. Initially $\mathcal{T}$
  contains only the root node. New nodes are generated from existing
  nodes, by selecting a node, \ie a generalized sequence, 
selecting one position where a $\Joker$
  appears, and generating $|\Alphabet|$
new nodes by replacing this $\Joker$ by all   possible letters 
$\alpha\in\Alphabet$. In this way
algorithm traverses the tree from the root towards the leafs
calculating an upper bound of the ground state energies of the sequences
represented by this node. 
Within the algorithm, two functions appear,
$\proc{Ground-State-Energy}(\Extended\Sequence_\alpha)$ and
$\proc{Ground-State-Bound}(\Extended\Sequence_\alpha)$, which 
 essentially use
\refeq{eq:nuss_recursion} to calculate the ground-state energy and the
upper bound for it, respectively. 
If this upper bound is below the energy
$E(\Structure)$ of the
structure \Structure, none of the sequences represented by this node has
this structure as a ground state, and the descend towards the children of this
node can be stopped here: the algorithm ignores this node by
not putting it into $\Tree$.
On the other hand, if a leaf node is reached and its ground state energy is
equal to the energy of the structure, a solution is found and the algorithm
terminates successfully.

The selection steps in line 4 and 5 require further explanations: We use a
stack-like data structure, so the last inserted sequence in line 
13 is used
first here (depth-first search).
 The selection step of a joker-letter in line 5 is more difficult: we tried
some strategies in which the next inserted base can be chosen. 
All this strategies were static ones, that means the order of insertion was
chosen based on the concrete structure given, but the order was fixed before
starting with the algorithm.
At the end we found the following strategy to be the best \footnote{%
  Surprisingly, we found that the insertion in plain order from 1 to $L$ is
  better than many other complicated strategies.
}:
We first insert paired
bases, and we choose the base pair $(i,j)$ first that encloses the most other
bases---\ie $\SubStructure{i,j}$ is the largest substructure of any
$(i,j)\in\Structure$. The procedure continues with the substructure
$\SubStructure{i+1,j-1}$, if it is not empty, or continues with a pair $(i',
j')\not\in\SubStructure{i+1,j-1}$ enclosing the next largest substructure. 
At the end we insert the unpaired bases.

 \subsubsection{Randomized steepest-descent Algorithm}
  \label{sec:randomized_algo}
  We further implemented a randomized algorithm for finding a solution of the
  design problem for a given structure \Structure\, similar to Ref. 
\onlinecite{SFS*94},
  while in Ref. \onlinecite{AFH*04} 
a much more sophisticated method is explained.
  We start with a compatible sequence, \eg every pair of the structure is
  assigned a \Letters A-\Letters U pair and all unpaired bases are assigned to
  \Letters G (again \Letters A if the  alphabet contains only two letters).  
  Either this already solves the design problem or we modify the sequence at
  one place as following: for the given sequence we calculate a ground-state
  structure $\Structure_0$, then we choose a pair $\Pair$, which is in exactly
  one of the structures \Structure\ and $\Structure_0$---\ie
  $\Pair\in\Structure\operatorname*\triangle\Structure_0$---and randomly
  modify one of this   two bases---if $\Pair\in\Structure$ we keep the other
  base complementary.  
  We accept this step, if the ground-state energy is not below of that of the
  previous sequence. 
  The procedure is repeated until a sequence is found that solves the
  design problem, or until a certain number of random steps has been executed,
  in this case, the algorithm stops unsuccessfully. 

  Of course, this method can never proof that a certain structure is
  undesignable. 
  However, we combined this strategy with the branch-and-bound algorithm
  above: whenever a rejection step takes place---\ie the condition in line
  14 of algorithm in \reffig{fig:branch_and_bound_algorithm}
  is reached---one random step with an independently stored sequence is
  done. 
  This can be quite efficient in the designable case, because on 
  average it 
  requires much less steps than the deterministic branch-and-bound
  algorithm.
On the other hand it
  doubles the efforts in the undesignable case. This pays off
    in particular for the
  four-letter case discussed in \refsec{sec:four_letter},
  because there almost all structures are designable.
  Especially, for design times much larger than the sequence length---\ie
  $\Designtime\gtrsim 10 L$---the random-method is almost always faster than
  the deterministic algorithm. 
  This is different in the two- and three-letter case, where the deterministic
  algorithm requires less steps.

\subsection{Generating random secondary structures}
\label{sec:structure_generation}
Later on we examine the designability of randomly generated secondary
structures for a given sequence length $L$. 
We parametrize our ensemble by the
probability $p$ that a certain base in the sequence is paired 
(for rRNA $p$ is typical in the range $0.6\dots0.8$ \cite{GYW05}).
We construct each sample in two steps: First,  
we draw the number of pairs $P$ of the structure from a binomial
distribution between 0 and $\lfloor L/2\rfloor$ centered at $p L/2$. Then, among all possible structures
of length $L$ having $P$ pairs, we select one randomly, 
such that each structure has
the same probability of being chosen. The achieve this, we have to perform a
preprocessing step first:

In the preprocessing step, 
we calculate the number $\SN(P, L)$ of possible structures of a sequence
of length $L$ and with $P$ pairs.
The number $\SN(P,L)$ is the number of 
possible structures $\SN(P,L-1)$ of the
smaller sequence plus the number of possible structures, where base
$L$ is paired with base $L-k$. Hence, the
 value $\SN(P,L)$ can be calculated by the following recursion
relation \cite{HSS98}: 
\begin{align}
\label{eq:structure_number_recursion}
&
\begin{aligned}
&\SN(P,L)= \SN(P, L-1)+\\
 &     \sum_{k=\hmin}^{L-1}\sum_{q=0}^{P-1} \SN(q,k-1)\,\SN(P-q-1, L-k-1),
\end{aligned}\\
&  \SN(P=0,L)=1,\quad \SN(P<0, L)=\SN(P>L/2,L)=0 \nonumber           
\end{align}
The first sum is over all possible distances between this two bases; the
second sum is over the number of pairs enclosed by the pair $(L-k, L)$.
The product is the number of possible structures having $q$ pairs enclosed by
$(L-k, L)$ and the remaining $P-q-1$ pairs in the range from $1$ to $L-k-1$. 
The construction of the matrix $\SN(P,L)$ requires $\mathcal{O}(L^4)$
calculation steps, but this is required only once for all lengths up to a
maximum length $L$.
Note that for 
 $\hmin=1$ the number of structures can be calculated explicitly 
$\SN(P,L)=\frac1{P+1} \binom{2P}{P} \binom{L}{2P}$.

\begin{figure}
\includegraphics*[width=\picturewidth]{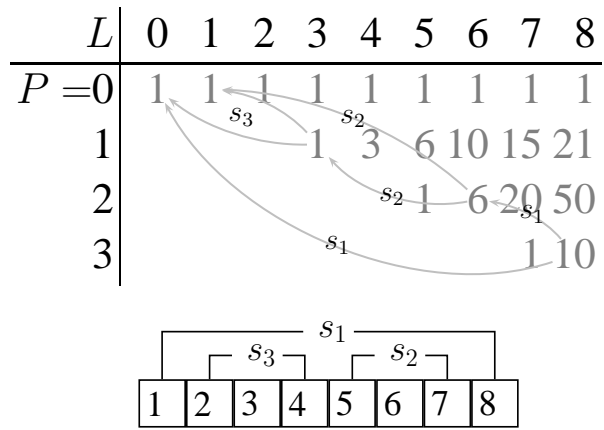}
  \caption{Example of the structure generation $\hmin=2$. Construction
    of a random 
    structure with $P=3$ and $L=8$. The way of construction a (random)
    structure from this, is indicated in the table by the arrows. 
    There are 10 possibilities to construct a structure of length 8 with 3
    pairs of bases. In step $s_1$ we choose to link base 1 and 8, which
    leaves a structure of length 6 with 2 pairs enclosed and a (trivial)
    structure of length 0 outside this pair. 
    In step $s_2$ we choose base 5 and 7 to be paired, leaving a trivial
    structure of length 1 enclosed and structure of length 3 with one pair
    outside. 
    For the latter there is only one choice, namely to connect base 2 and 4
    (step $s_3$). 
    The resulting structure is shown in the figure.   
  \label{tab:catalan_numbers}
  }
\end{figure}

Now, for each sample to be generated, where the number $P$ of pairs has
  been randomly chosen as explained above, the actual structure is
  selected in the following way. First, note that
depending on $\hmin$ there are values of $P$ and $L$, where 
no  structures exist, \ie $\SN(P,L)=0$, these
cases are rejected immediately.
Otherwise,
the random structure is constructed with a backtracing algorithm: starting at
$\SN(P,L)$ choose one of the summands according to its weight, insert the
corresponding pair to the structure and recurs into the sub-sequences.
As an example %
we show the random generation of a structure of length $L=8$
with $P=3$ pairs (see \reffig{tab:catalan_numbers}). 
The non-zero contributions to
$\SN(3,8)=\SN(3,7)+\SN(0,1)\SN(2,5)+\SN(1,3)\SN(1,3)+\SN(2,5)\SN(0,1)+\SN(2,6)\SN(0,0)$, 
each of the summands represents a possible pairing of base number 8 with
another base---with the exception of the first summand, which counts the
number of possible structure, where base number 8 is not paired at all.
We choose the last summand, meaning that base 8 is paired with base 1. 
Leaving two pairs which must be distributed between the bases from 2 to
7; here we choose to pair base 7 with base 5, leaving only one possibility
for the remaining pair, namely base 4 paired with base 2.

Finally, note that the
 average number of structures available for  given $p$ and $L$
is given by
    \begin{equation}
      \label{eq:average_structure_number}
      \aSN(p,L)=\sum_{P=0}^{\lfloor L/2\rfloor} \binom{\lfloor L/2\rfloor}{P}\, p^{P}\,
      (1-p)^{\lfloor L/2\rfloor-P} \SN(P,L)\,.
    \end{equation}

\section{Numerical Results}
\label{sec:results}
For an ensemble of randomly chosen structures of given sequence
length $L$ we examined, whether these structures are designable or not. 
We used different alphabets with two, three and four letters. 
All calculations for the results presented below 
were performed with the parameters
$\PairE=-2$, $\StackE=-1$, and $\hmin=2$. Note that 
increasing the stacking energy
\StackE\ in comparison to the pair energy makes the design problem more
difficult: in the limit $\StackE\to-\infty$ it would be favorable to remove
all non-stacked pairs from the structure, if this allows only one additional
stacked pair. 
Considering the minimum distance \hmin\ between two paired bases of natural
RNA, it seems to be more appropriated to use a larger value for \hmin,
\eg $\hmin=5$ would be more appropriate, but
this increases the computational effort without changing the qualitative
results: only $\hmin=1$ has a different qualitative behavior (see
\reffig{fig:hmin_counter_example}).

\subsection{Two-letter alphabet}
\label{sec:two_letter}
The alphabet consists of two complementary letter, \eg \Letters A and \Letters  U, only. 
\begin{figure}[htbp]
  \includegraphics*[width=\picturewidth]{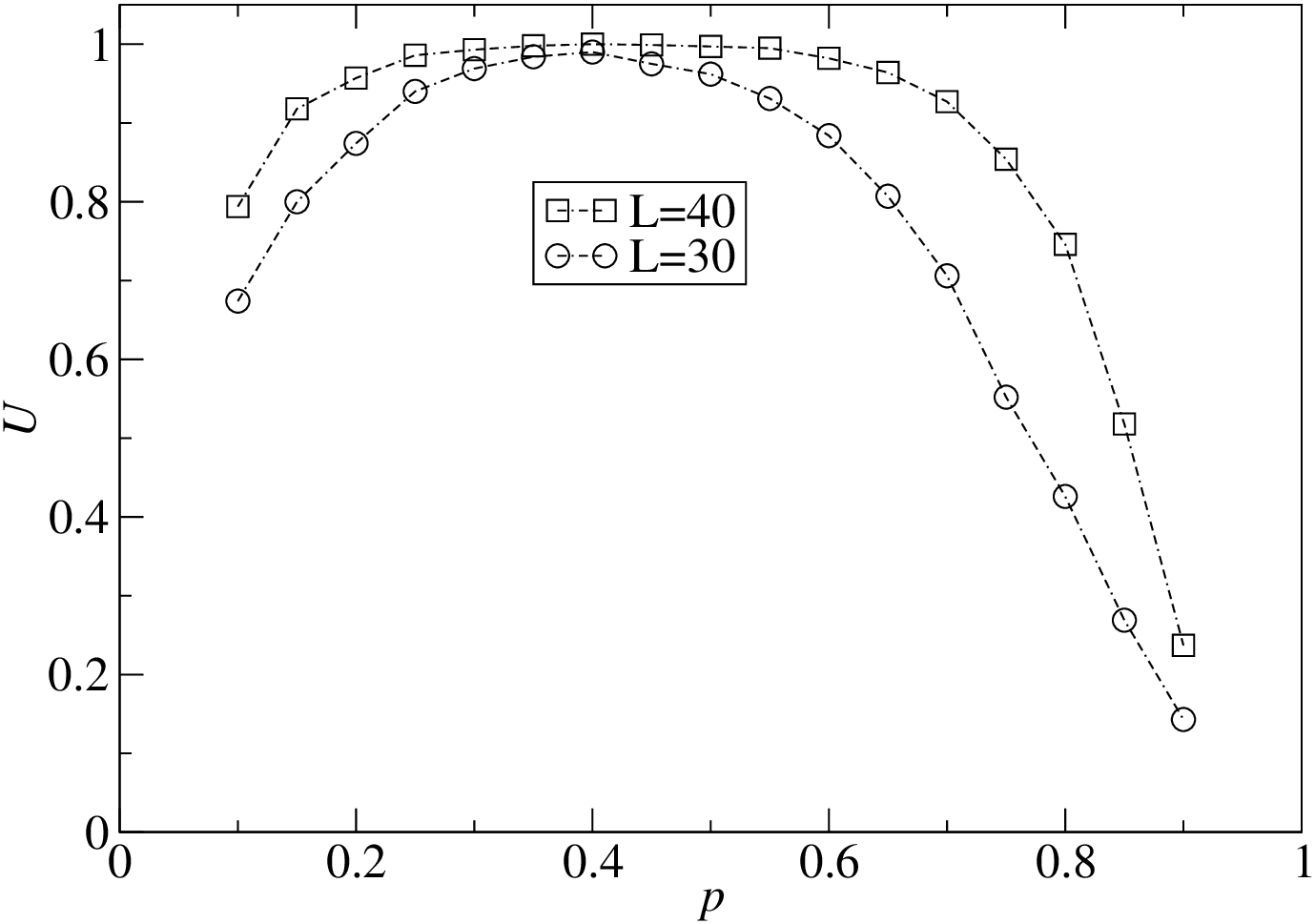}%
  \caption{%
    The undesignability \UnDesignability\ of random structures for an
    underlying two-letter alphabet is shown as function of the probability
    $p$ that a base is paired. 
    Even for small sequences and low probabilities of bases being paired,
    almost all structures are undesignable. 
    \errorbars
    \legendlines
  \label{fig:two_letter}
  }  
\end{figure}
In  \reffig{fig:two_letter} the fraction \UnDesignability\ of the
  undesignable structures is shown as a function of the probability $p$ that a
  base is paired.
  For small $p$ the fraction \UnDesignability\ for all lengths $L$ increases
  quickly with growing $p$ 
from small values to its maximum possible value close to
one. Thus, in particular for moderate RNA lengths $L\approx 100$,
  almost no structure is designable. 
   For structures, where many bases are paired, only a quite restricted
  class of structures is possible, \ie 
structures with many nested base-pairs,
  which have obviously a high probability to be designable. 
  For this reason the undesignability $U$ decreases
 again for larger $p$.

  For fixed $p$-values the value of
\UnDesignability\ increases with the sequence length
  $L$, which seems to be plausible because, if a structure of small length is
  undesignable, larger structure containing this structure must be also
  undesignable. 

  We conclude that two letters do not suffice to provide a large variety of
  secondary structures needed in nature to perform the large
  number of required RNA functions.

\subsection{Three-letter alphabet}
\label{sec:three_letter}
\begin{figure}[htbp]
  \includegraphics*[width=\picturewidth]{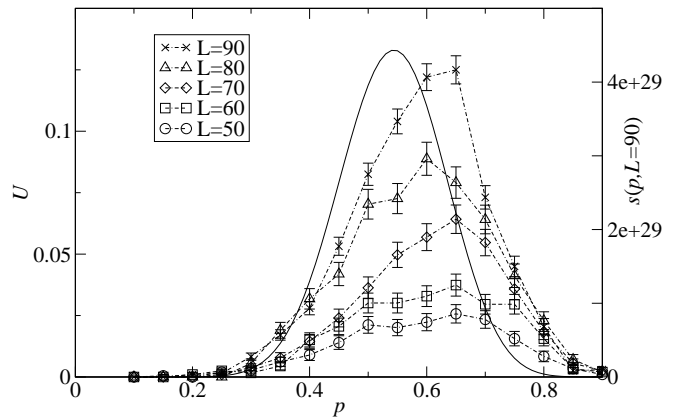}%
  \caption{The undesignability \UnDesignability\ of random structures for an
    underlying three-letter alphabet is shown as function of the probability
    $p$ that a base is paired.  
    In comparison to the two-letter case (\reffig{fig:two_letter}) 
    many more  of structures are designable, but still a reasonable fraction of
    structures is undesignable. 
    In light gray 
    the average number $\aSN(p, L=90)$ of structures of length
    90 is shown (see \refeq{eq:average_structure_number}): The maximum of this
    curve is at smaller $p$-value than the maximum of $\UnDesignability(p,L=90)$. 
    \legendlines
  \label{fig:three_letter_dsgnability}
}
\end{figure}

The alphabet consists of two complementary letter, \eg \Letters A and \Letters U, and one
additional letter, \eg \Letters C, not complementary to any other letter. 
As one can see from \reffig{fig:three_letter_dsgnability} compared to the two letter case
a larger amount of structures is designable, but with larger sequence lengths
still a larger fraction becomes undesignable. 

\begin{figure}[htbp]
  \includegraphics*[width=\picturewidth]{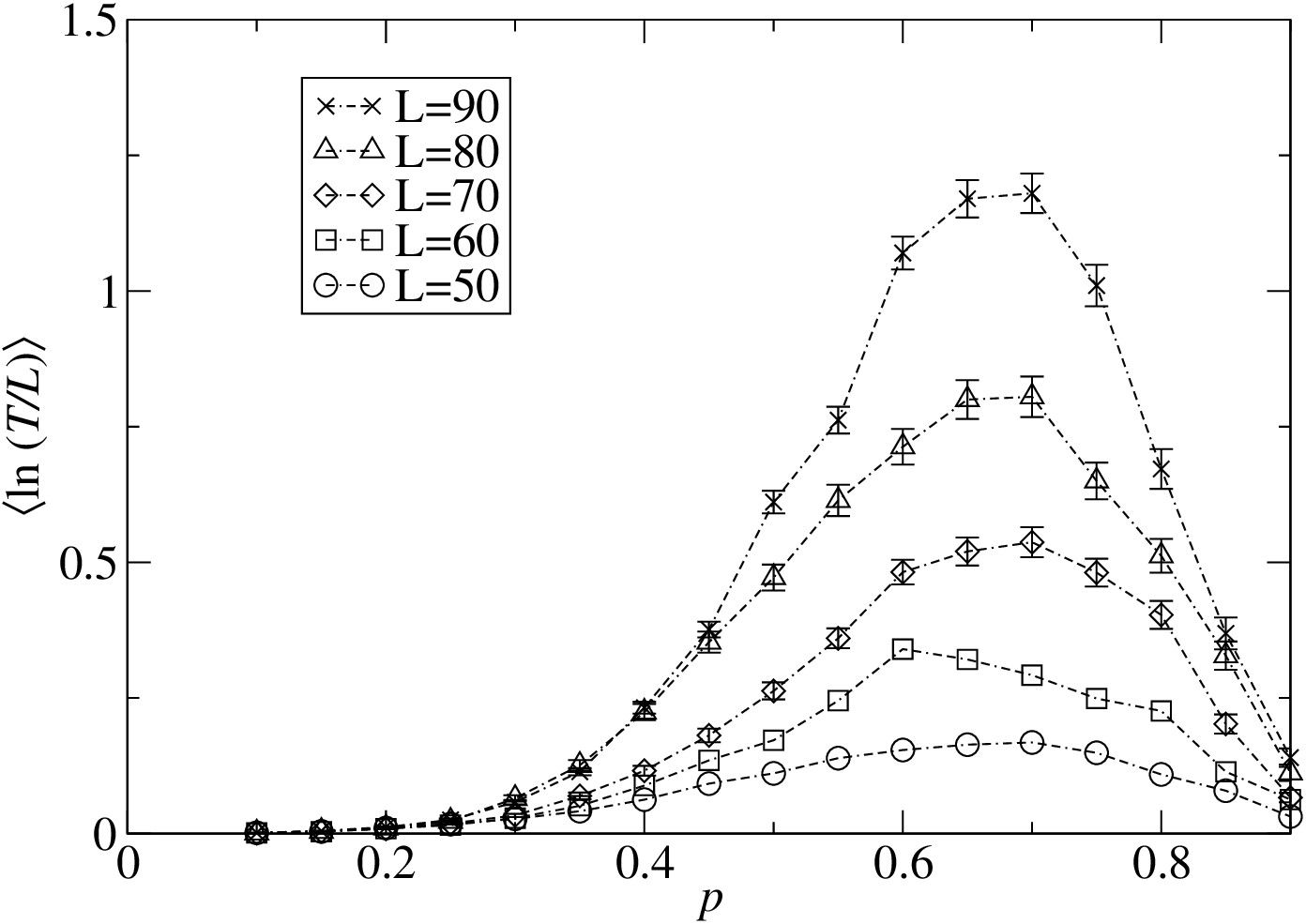}%
  \caption{For the three-letter alphabet the design time \Designtime\ for
    designable structures is show as a function of the pairing probability
    $p$. 
    The positions of the maxima are at similar positions as the corresponding
    maxima in  \reffig{fig:three_letter_dsgnability}.
    \errorbars
    \legendlines
  \label{fig:three_letter_dsgn_time}
  }  
\end{figure}

We also looked at the ``time'' $T$ required to find a solution---if any
exists. 
``time'' means here, how often either of the two functions
$\proc{Ground-State-Energy}(\Extended\Sequence_\alpha)$ or 
$\proc{Ground-State-Bound}(\Extended\Sequence_\alpha)$ (see
\reffig{fig:branch_and_bound_algorithm}) is called; because this two function
are called at least $L$-times, $T$ is at least $L$. 
In \reffig{fig:three_letter_dsgn_time} the average of  $\ln(T/L)$ is shown as
a function of $p$. 
Because $T\geq L$ a value close to zero indicates, that a solution is
found (on the average) almost immediately.

\begin{figure}[htbp]
  \includegraphics*[width=\picturewidth]{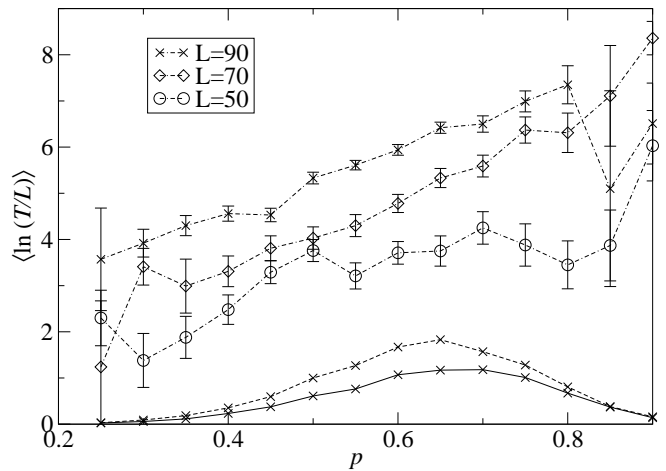}%
  \caption{
    The average time for the undesignable structures required to verify the
    undesignability  is shown for three different lengths (black lines) for
    the three-letter case. 
    For comparison in the lower part of the figure the average time
    for designable structures (solid curve; cf.\
      \reffig{fig:three_letter_dsgn_time}) and the average time to proof 
    either designability  or undesignability (dashed  curve) are shown.
    In general the higher the pair probability $p$ is the more difficult it
    becomes to proof the undesignability. 
    Further one can see that it is much more difficult to prove
    undesignability than to find a solution in the designable case. 
    For probabilities $p$ below $0.3$ or above $0.8$ only few structures are
    undesignable and the corresponding error bars become large---\ie more
    samples are required to get better results in this regime.
    \legendlines
  \label{fig:three_letter_undsgn_time}
  }  
\end{figure}

The maxima of this curves are almost at the positions as that of
\reffig{fig:three_letter_dsgnability}, meaning that for values of $p$, where a
large fraction of structure is undesignable, it is difficult---\ie requires
many steps---to find a solution for the designable structures.
The structures which are not designable behave a bit
  differently, cf.\ \reffig{fig:three_letter_undsgn_time}. There the
  time needed to prove that no design is possible increases
  monotonously with $p$, and is much larger than the time needed to
  find a solution in the designable cases. 
Nevertheless, the total running time of the
  branch-and-bound algorithm is mostly determined by the designable
  case, hence we observe a peak close to $p=0.6$ as well, see lower
  curve in \reffig{fig:three_letter_undsgn_time}.
This behavior of the running time 
is similar to the behavior found for suitable random ensembles  
of classical combinatorial
  optimization problems \cite{HW05a,GJ79}%
, as observed for 
the satisfiability problem \cite{MSL92} %
or
the vertex-cover problem \cite{WH00}%
. 
Also in these and other cases, the running
time of exact algorithms similar to branch-and-bound   increases strongly
when the average number of unsolvable random instances increases. The
only difference to the present case is that for these classical
optimization problems in the limit of diverging system sizes,
phase transitions between solvable and
unsolvable phases can be observed. In the case of RNA secondary
structures, we are interested only in finite lengths, because in nature
finite (rather short) RNA sequences dominate anyway. 

Finally,
we also want to mention that the maximum of the average number of structures
$\aSN(p,L)$---as shown in \reffig{fig:three_letter_dsgnability}---is at a
slightly smaller $p\approx0.54$ than the maxima of $\UnDesignability(p)$ and
$\average{\ln T/L}$. Hence, in contrast to the two-letter
  case, there is at least window of $p$ values, where a large number
  of designable structures exist. On the other hand, in the range
  $p\in [0.6\ldots 0.8]$, where most of the wild-type RNA can be
  found, the number of designable structures is still small. 
  Especially, for sequence lengths $L\gtrsim1000$ we expect that
    again most structures are undesignable.
  Hence,
  three letters seem also not to be sufficient.

\subsection{Four-letter alphabet}
\label{sec:four_letter}
The alphabet consists of two pairs of complementary letters, \eg
\Letters A, \Letters  U and \Letters C, \Letters G. 
In this case we observe that for all lengths up to $L=90$ the
undesignability \UnDesignability\  is essentially zero---\ie so far we
have not found any random structure  that is undesignable. 
This means that four letters are sufficient, at least for
  moderate system lengths, to design all possible structures maybe
  needed in cell processes. Nevertheless, 
as shown in \refsec{sec:discussion_designability} structures exist, that are
undesignable even in the four-letter case, but such
structures must be quite rare for lengths up to $L=90$. This
  means that in the limit of infinite RNA lengths, which is only of
  abstract academic interest, almost all \emph{random}
  structures become undesignable, because the probability that somewhere
in the infinite sequence there is an undesignable subsequence of
finite  length is one, as explained in the next section. 
Since, as already pointed out
above, naturally occurring RNA have to be only of rather restricted length to
perform their functions, this
effect has no influence and a four letter alphabet seems to be sufficient.

\begin{figure}[htbp]
  \includegraphics*[width=\picturewidth]{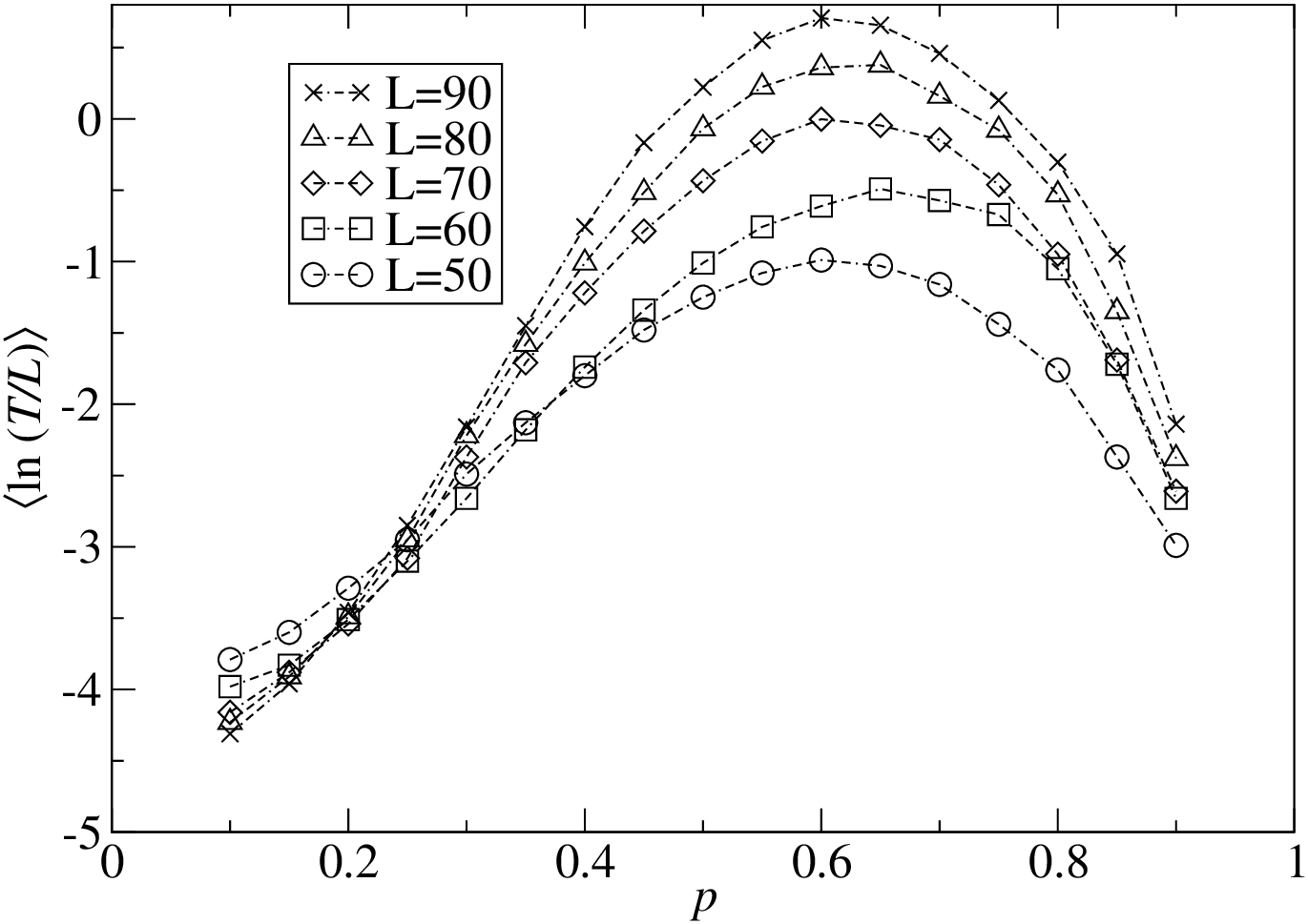}%
  \caption{For the four-letter alphabet the design time \Designtime\ for
    designable structures is shown as a function of the pairing probability
    $p$. 
    The positions of the maxima are at similar positions as the corresponding
    maxima in  \reffig{fig:three_letter_dsgn_time}.
    \errorbars
    \legendlines
  }  
  \label{fig:fourletterdsgntime}
\end{figure}

In \reffig{fig:fourletterdsgntime} we  show the average
``time'' \Designtime\ to 
find a solution as a function of $p$, but here we used a combined
deterministic-randomized algorithm, which is quite fast for low pairing
probabilities---\ie $p<0.4$---where on the average less than $L$ ground-state
calculations are necessary to find a solution. 
On the other hand for values $p\approx 0.6$  the design time \Designtime\
seems to grow faster than exponentially in the sequence length $L$.
This strong increase of the running time is \emph{not}
  accompanied by an increase of the undesignability \UnDesignability\,
at least not on the length scales we can access with the algorithm,
since we do not find any undesignable structures in this range. 
This is different from
the three-letter case and from the classical optimization problems
cited above. Nevertheless, it is striking that the structures which
are hardest to design are close to the region $p\in[0.6\ldots 0.8]$,
where the naturally occurring RNA secondary structures can be
found. Furthermore, this strong increase of the running times means
that one \emph{cannot} use the randomized algorithm to look quickly for
probably undesignable structures in the four-letter case: One cannot
just stop searching after a search time which only increases
polynomially with the sequence lengths, because in this case one
would even miss the designable structures. Hence, longer RNA, \ie
random RNA which are not designable, seems out
of reach currently.

\subsection{Discussion}
\label{sec:discussion_designability}
While in the two letter case a large amount of random structures is not
designable, only a small amount of them is undesignable when using a
three-letter alphabet. 
In the four-letter case designability seems to dominate the structure space by
far: in fact, so far we have not found any random structure which is
undesignable for the given parameter ($\StackE=-1,\,\PairE=-2,\,\hmin=2$).
This leads to the question, whether there are any undesignable structures at
all. 

\begin{figure}[htbp]
 \includegraphics*[width=\picturewidth]{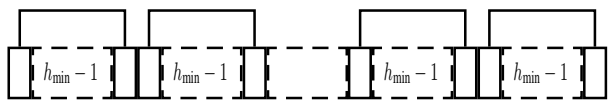}
  \caption{Principle of a non-designable structure. 
    Structures consisting of a repeated pattern of simple paired bases become
    undesignable, if this pattern is repeated often enough. For result see
    \reftab{tab:undesignable_structures}. 
  \label{fig:non_solution_structure}
  }  
\end{figure}

Indeed, there are such structures (see \reffig{fig:non_solution_structure}):
 for a given length $L$ build a non-nested
structure by the pairs $((\hmin+1) n+1,\,(\hmin+1)(n+1))$ with
$n=0,1,2,\dots$ and $(\hmin+1) (n+1)\leq L$.
  Such structures are a examples for chains: a \emph{chain} \Chain\ of length
  $l$ is a set of pairs $\Chain=\{(i_1,j_1),\,(i_2, j_2),\dots, (i_l, j_l)\}$  
  with the property $j_n+1=i_{n+1}$ for $n=1,\dots,l-1$.
  A chain \Chain\ which is a subset of a structure \Structure, \ie
  $\Chain\subset\Structure$, is called a \emph{subchain} of \Structure. 
Chains of large enough lengths, \eg the structure sketched in
\reffig{fig:non_solution_structure}, are undesignable
 for a similar reason as the structure
shown in \reffig{fig:hmin_counter_example} is undesignable (with $\hmin=1$):
there are only finite many possible combinations of bases
being paired, such that after a while a repetition
  occurs. Nevertheless, the argument is more
complex here and we do not go into details. We only show
in \reftab{tab:undesignable_structures} the minimum length of structures
sketched in \reffig{fig:non_solution_structure} for which these
become undesignable for different \hmin\ and the corresponding
  running times of the branch-and-bound algorithm.

\begin{table}[htbp]
  \centering
  \begin{tabular}{*{4}{|>{$}c<{$}}|}
    \hline
    \hmin & L & \text{pairs} & \Designtime\\
    \hline
    2 & 48 & 16 & 6\cdot10^{7}\\
    3 & 60 & 15 & 5\cdot10^{8}\\
    4 & 75 & 15 & 2\cdot10^{9}\\
    \hline
  \end{tabular}
  \caption{The minimum length of structures according to
    \reffig{fig:non_solution_structure} that are undesignable. 
    In the last column the time \Designtime\ required to prove the undesignability
    with the brand-and-bound algorithm is shown. 
  \label{tab:undesignable_structures}
}
\end{table}

This implies that structures \Structure\ which contain a subchain \Chain\ of
length $l\geq16$ are also undesignable. 
In the limit $L\to\infty$ with pair probability $p>0$ we expect that almost all random
structures contain a subchain of size $l\geq16$, thus making this structures
undesignable.   
However, for native RNA this limit is not relevant:
For an ensemble of $10.000$ random structures of length $L=1024$ and pair
probability $p=0.7$ we looked for each structure for the subchain of the
longest length $l$ and found none longer than 11.
Assuming that all undesignable structures in the four-letter case are
undesignable because they contain a subchain longer than $l=15$, such
structures are very rare even for biological lengths.

Finally, we shortly want to mention the five-letter case: 
two pairs of two
complementary bases (\Letters{A}-\Letters{U}, \Letters{C}-\Letters{G}) and an
unpairable fifth letter (\eg \Letters{X}).  
In this case it is easy to see that even structures as explained in
\reffig{fig:non_solution_structure}  are designable: 
Start with a sequence of type \Letters{ACUGACUGACUGACUG\dots}, replace the
bases at positions 2,5,8,\dots with $\hmin-1$ letters of type \Letters{X}, \eg
yielding in the case $\hmin=2$: \Letters{AXUGXCUXACXGAXUG\dots}.
First, in this sequence stacked-pairs are impossible, because for non pair
$\Base[i]\Base[i+1]$ there is a required complementary pair
$\cBase[i+1]\cBase[i]$.  Further, this sequence is compatible to the
structure and there are 
exactly as many complementary bases pairs as there are pairs in the structure.
Of course, this does not prove that with five letters all structures are
designable, but undesignable structures are at least expected to be
even much less frequently than in the four-letter case.

\section{Summary}
\label{sec:summary}
We numerically investigated the RNA secondary structure design problem for
different alphabet sizes. 
We used a deterministic branch-and-bound algorithm to get definite answers,
whether a given structure is designable or not.
Due to efficiency reasons in the designable cases, we combined this algorithm
with an probabilistic one, gaining significantly performance improvements in
the four-letter case. 

We examined the designability for an ensemble of random structures as a
function of the probability that a base of sequence is paired. 
Our findings for the two-letter case are that it is almost impossible to design
most of the structures.
In the three-letter case already for small sequence sizes ($L\approx 90$)
about 10\% of the structures are undesignable for biological relevant pairing
probabilities, leading to the conclusion that for biological sequence sizes
($L\approx 1000$) again most structures are undesignable. 

Interestingly, this changes when going to the (natural) four-letter alphabet: 
within our studies we have not found a single random structure that we could
prove to be undesignable.
Although, there are structures that are undesignable, they occur with very
low frequencies. 

We further studied the computational 
time required to design a structure. 
Although, this for sure depends strongly on the algorithm, we found in
three-letter case that required time is maximal in the regime where the
undesignability is largest.
In the four-letter case the design times look similar to that of the
three-letter case: again we observed 
a maximum of the design times in for
$p\approx0.6$, close to the region where naturally occurring
  RNA can be found. Although, (almost) all structures are designable, it is
sometimes difficult to design them.

\begin{acknowledgments}
The authors have obtained financial support from the
\emph{Volks\-wagen\-stif\-tung} (Germany) within the program
``Nachwuchsgruppen an Universit\"aten''.
The simulations were performed at the Paderborn Center
for Parallel Computing in Germany and on a workstation
cluster at the Institut f\"ur Theoretische Physik, Universit\"at
G\"ottingen, Germany.
We thank M.~Jungsbluth for helpful remarks.
\end{acknowledgments}

\end{document}